# RECENT BEAM–BEAM EFFECTS AT VEPP-2000 AND VEPP-4M*

D.B. Shwartz, D.E. Berkaev, I.A. Koop, E.A. Perevedentsev,
Yu.A. Rogovsky, Yu.M. Shatunov, Budker Institute of Nuclear Physics, Novosibirsk State University, Novosibirsk, Russia
E.B. Levichev, A.L. Romanov, D.N. Shatilov, P.Yu. Shatunov, Budker Institute of Nuclear Physics, Novosibirsk, Russia

*Abstract*

Budker INP hosts two $e^+e^-$ colliders, VEPP-4M operating in the beam energy range of 1–5.5 GeV and the low-energy machine VEPP-2000, collecting data at 160–1000 MeV per beam. The latter uses a novel concept of round colliding beams. The paper presents an overview of observed beam–beam effects and obtained luminosities.

## VEPP-4M

Being a rather old machine with a moderate luminosity, VEPP-4M has several unique features, firstly a very low beam-energy spread, and a system for precise energy measurement, providing an interesting particle physics program for the KEDR detector. Over recent years VEPP-4M was taking data at a low energy range with two bunches in each beam. The luminosity at this range is limited by beam–beam effects with the threshold beam–beam parameter $\xi_y \leq 0.04$ [1]. In this case the luminosity depends on energy as $L \propto \gamma^4$ (see Fig. 1).

The main parameters of the VEPP-4M collider are listed in Table 1.

## ROUND COLLIDING BEAMS

The VEPP-2000 collider [2] exploits the round beam concept (RBC) [3]. The idea of round-beam collisions was proposed more than 20 years ago for the Novosibirsk Phi-factory design [4]. This approach, in addition to the geometrical factor gain, should yield the beam–beam limit enhancement. An axial symmetry of the counter-beam force together with the $X$–$Y$ symmetry of the transfer matrix between the two IPs provide an additional integral of motion, namely, the longitudinal component of angular momentum $M_z = x'y - xy'$. Although the particles' dynamics remain strongly nonlinear due to beam–beam interaction, it becomes effectively one-dimensional. Thus there are several demands upon the storage ring lattice suitable for the RBC:

___________________________________________

*Work supported by the Ministry of Education and Science of the Russian Federation, grant N

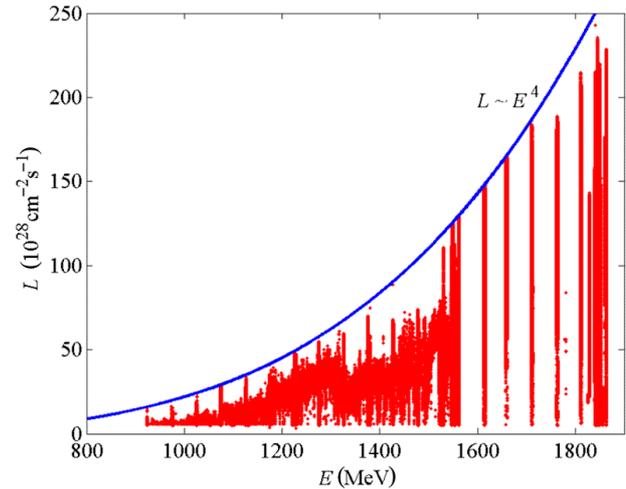

Figure 1: VEPP-4M luminosity dependence on beam energy.

Table 1: VEPP-4M main parameters.

| Parameter | Value |
|---|---|
| Circumference ($C$) | 366 m |
| Energy range ($E$) | 1–5.5 GeV |
| Number of bunches | 2 × 2 |
| Betas and dispersion at IP ($\beta^*_x$, $\beta^*_y$, $\eta^*$) | 75, 5, 80 cm |
| Betatron tunes ($\nu_{x,y}$) | 8.54, 7.57 |
| Beam–beam parameters ($\xi_x$, $\xi_y$) | 0.025, 0.04 |
| Luminosity at 1.85 GeV ($L$) | $2.3 \times 10^{30}$ cm$^{-2}$ s$^{-1}$ |

i) head-on collisions (zero crossing angle);

ii) small and equal $\beta$ functions at IP ($\beta^*_x = \beta^*_y$);

iii) equal beam emittances ($\varepsilon_x = \varepsilon_y$);

iv) equal fractional parts of betatron tunes ($\nu_x = \nu_x$).

The first three requirements provide the axial symmetry of collisions while requirements (ii) and (iv) are needed for *X–Y* symmetry preservation between the IPs.

A series of beam–beam simulations in the weak–strong [5] and strong–strong [6] regimes were done. Simulations showed the achievable values of beam–beam parameters as large as $\xi \sim 0.15$ without any significant blow-up of the beam emittances.

## VEPP-2000 OVERVIEW

The layout of the VEPP-2000 complex is presented in Fig. 2. The complex consists of the injection chain (including the old beam production system and Booster of Electrons and Positrons (BEP) with an energy limit of 800 MeV) and the collider itself with two particle detectors, Spherical Neutral Detector (SND) and Cryogenic Magnetic Detector (CMD-3), placed into dispersion-free low-beta straights. The final focusing is realized using superconducting 13 T solenoids. The main design collider parameters are listed in Table 2.

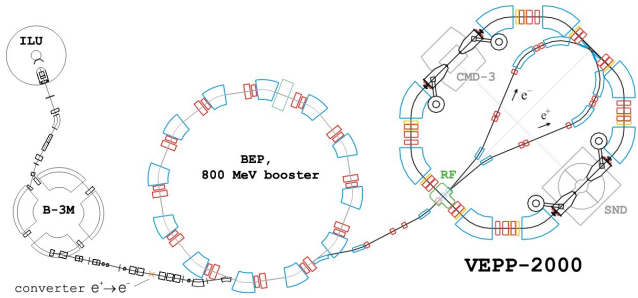

Figure 2: VEPP-2000 complex layout.

Table 2: VEPP-2000 main parameters (at $E$ = 1 GeV).

| Parameter | Value |
| --- | --- |
| Circumference ($C$) | 24.3883 m |
| Energy range ($E$) | 200–1000 MeV |
| Number of bunches | $1 \times 1$ |
| Number of particles per bunch ($N$) | $1 \times 10^{11}$ |
| Betatron functions at IP ($\beta^*_{x,y}$) | 8.5 cm |
| Betatron tunes ($\nu_{x,y}$) | 4.1, 2.1 |
| Beam emittance ($\varepsilon_{x,y}$) | $1.4 \times 10^{-7}$ m rad |
| Beam–beam parameters ($\xi_{x,z}$) | 0.1 |
| Luminosity ($L$) | $1 \times 10^{32}$ cm$^{-2}$ s$^{-1}$ |

The density of magnet system and detectors components is so high that it is impossible to arrange a beam separation in the arcs. As a result, only a one-by-one bunch collision mode is allowed at VEPP-2000.

## BEAM DIAGNOSTICS

Beam diagnostics is based on 16 optical CCD cameras that register the visible part of synchrotron light from either end of the bending magnets and give full information about beam positions, intensities, and profiles (see Fig. 3). In addition to optical beam position monitors (BPM), there are also four pick-up stations in the technical straight sections, two photomultipliers for beam current measurements via the synchrotron light intensity, and one beam current transformer as an absolute current monitor.

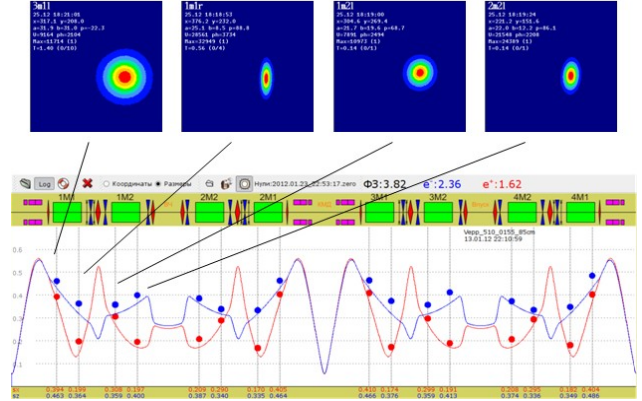

Figure 3: Beam profile measurements.

## CIRCULAR MODE OPTIONS

The RBC at VEPP-2000 was implemented by placing two pairs of superconducting focusing solenoids into two interaction regions (IR) symmetrically with respect to collision points. There are several combinations of solenoid polarities that satisfy the round beams' requirements: 'normal round' (++ −−), 'Möbius' (M) (+ + −+) and 'double Möbius' (DM) (++ ++) options rotate the betatron oscillation plane by ±90° and give alternating horizontal orientation of the normal betatron modes outside the solenoid insertions.

Two 'flat' combinations (+− +− or +− −+) are more simple and also satisfy the RBC approach if the betatron tunes lie on the coupling resonance $\nu_1 - \nu_2 = 2$ to provide equal emittances via eigenmodes coupling.

All combinations are equivalent in focusing and give the same lattice functions. But the tunes for M and DM options are different due to additional clockwise and counter-clockwise circular mode rotations (see Fig. 4).

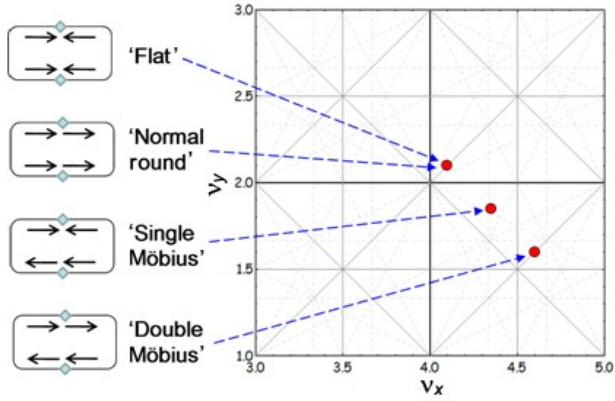

Figure 4: VEPP-2000 round beam options.

Unfortunately, computer simulations showed a serious limitation of the dynamic aperture (DA) for options with mode rotations. A brief experimental study was carried out upon the DM option. At first glance, this case could be preferable, because the tune is a little above 0.5 instead of an integer for the 'flat' mode. However, both the simulation and measurement gave a DA of only ~10 $\sigma_{x,y}$. Such studies should probably be continued for other options.

## LINEAR CONSIDERATIONS

An important feature of the VEPP-2000 lattice is the strong dependence of radiative emittance on the value of $\beta^*$. The decrease of $\beta^*$ causes emittance growth in such a way that $\sigma^{*2} = \beta^* \varepsilon = \mathrm{inv}(\beta^*)$. The expression for luminosity can be written in this case as

$$L = \frac{N^- N^+ f_0}{4\pi\sigma^{*2}} = \frac{4\pi\gamma^2 \xi^2 \sigma^{*2} f_0}{r_e^2 \beta^{*2}} \quad . \qquad (1)$$

One can now see that, although the specific luminosity does not depend on the choice of the value of $\beta^*$, the maximum luminosity limited by the beam–beam interaction with a given threshold $\xi_{th}$ can be higher for a lower $\beta^*$. The $\beta^*$ once optimized for a given aperture value at the top energy of 1 GeV should be decreased for lower energies corresponding to smaller radiative emittance to minimize the luminosity roll-off. Instead of ($\beta^*$ = const, $\varepsilon \propto \gamma^2$, $\sigma^* \propto \gamma$, $L \propto \gamma^4$), the energy scaling can be done as ($\beta^* \propto \gamma$, $\varepsilon \propto \gamma$, $\sigma^* \propto \gamma$, $L \propto \gamma^2$) (see the dashed blue and solid red lines in Fig. 7, respectively). Of course, this approach is very optimistic since it does not take into account the intrabeam scattering (IBS) emittance growth at a low energy as well as DA problems for a low $\beta^*$.

Similarly to the variation of $\beta^*$ caused by lattice tuning, the linear beam–beam simulation as well as weak–strong beam–beam simulations (LIFETRAC software program [7]) predict the inverse variation of the dynamic beta and dynamic emittance so that the beam sizes at IP are left unchanged by the linear beam–beam effect. At the same time, the size of the beam at the profile monitors around the ring varies strongly with the counter beam current (see Fig. 5).

## LUMINOSITY MEASUREMENTS

At VEPP-2000 luminosity monitoring is available from both detectors. Electrons and positrons from elastic scattering are easily detected in coincidence by the detector's calorimeters with an efficiency near 100% and counting rates of about 1 kHz at $L = 1 \times 10^{31}$ cm$^{-2}$ s$^{-1}$.

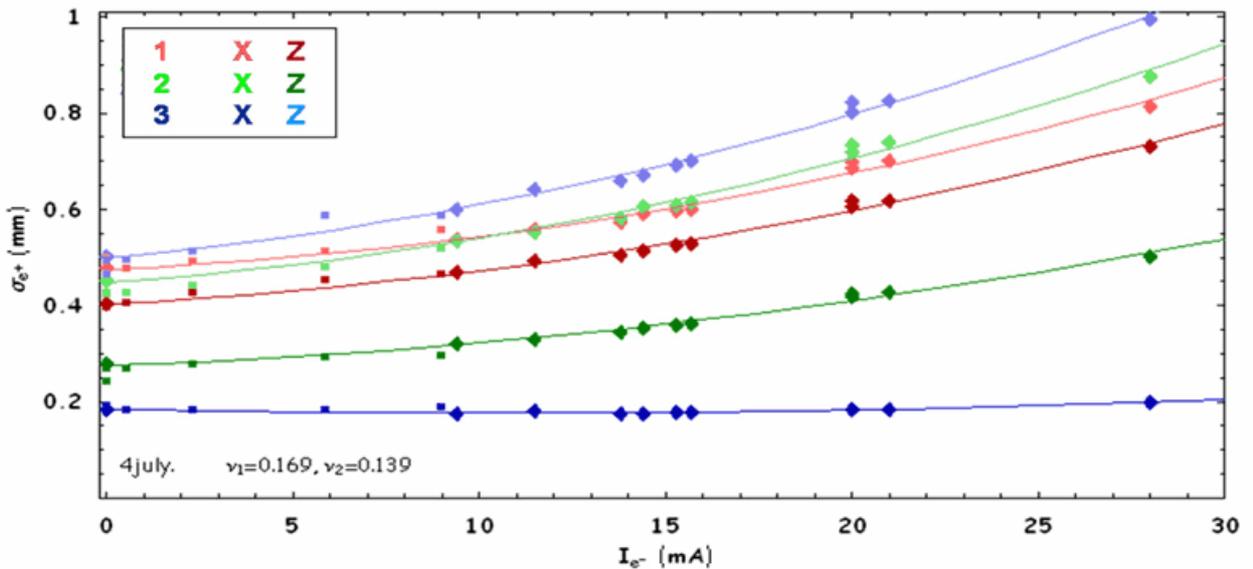

Figure 5: Weak–strong test of beam sizes growth with the counter beam current.

For prompt collision tuning a method for making luminosity measurements was developed based on the beam size data from the optical diagnostics. To calculate the luminosity one need know only the beam currents and sizes at the IP. As discussed above, due to the beam–beam effects the lattice functions and beam emittances show a significant current-dependent difference from their design values.

Assuming no focusing perturbations in the lattice other than those caused by the collision, and thus located at the IP, one can use transport matrices verified by the arc optics model to evaluate the beam sizes at the IP from the beam size measurements made by CCD cameras placed around the ring. Eight measurements for each betatron mode of the both beams are more than enough to evaluate the dynamic beta functions and dynamic emittances of the modes.

The accuracy of the method degrades at high beam intensities close to beam–beam threshold, where the beam distribution deviates from the Gaussian. Data from this luminometer, taken regularly during two hours at an energy $E = 800$ MeV, is presented in Fig. 6.

The advantages of this technique over the SND and CMD-3 luminosity monitors are the higher measurement speed and lower statistical jitter. The accuracy of the new method is nominally about 3–4% and it does not depend on the luminosity level, in contrast to the detector's data. On the other hand, the new technique is not sensitive to any possible focusing difference in two IPs. Generally, those three monitors give results coinciding within 10% accuracy.

## EXPERIMENTAL RUNS

VEPP-2000 started data-taking with both detectors installed in 2009 [8]. The first runs were dedicated to experiments in the high-energy range, while during the last 2012 to 2013 run an energy scan to the lowest energy limit was done. Apart from partial integrability in beam–beam interaction the RBC gives a significant benefit in the Touschek lifetime when compared to traditional flat beams. This results in the ability of VEPP-2000 to operate at an energy as low as 160 MeV — the lowest energy ever obtained in $e^+e^-$ colliders. The luminosity obtained during the last three seasons is shown in Fig. 7 with olive, magenta, and blue points. The red line is a naive estimate of the maximum achievable peak luminosity (jumps correspond to different commutation of the solenoid coils available at low energy). The blue dashed line shows the beam–beam limited luminosity for a fixed machine lattice. Black triangles and squares depict the peak and average luminosity achieved by the previous collider VEPP-2M [9]. Black circles indicate VEPP-2M luminosity without the superconducting wiggler.

For different energies the luminosity is limited for different reasons. At high energies (>500 MeV) it is limited mostly by an insufficient positron production rate. At energies over 800 MeV the necessity of energy ramping in the collider storage ring additionally restricts the luminosity. For lower energies the luminosity is limited by the beam–beam effects, especially by the flip-flop effect (see below). At the lowest energies the main limiting factors are the small DA, IBS, and low beam lifetime.

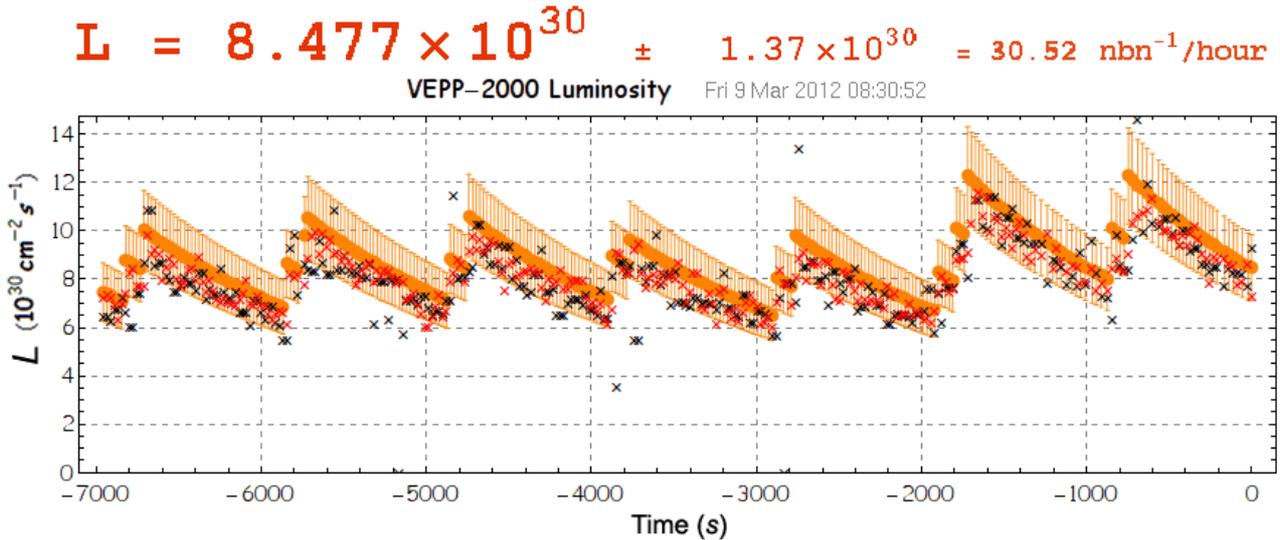

Figure 6: Luminosity at the energy $E = 800$ MeV. Black and red crosses, detectors; orange dots, luminometer.

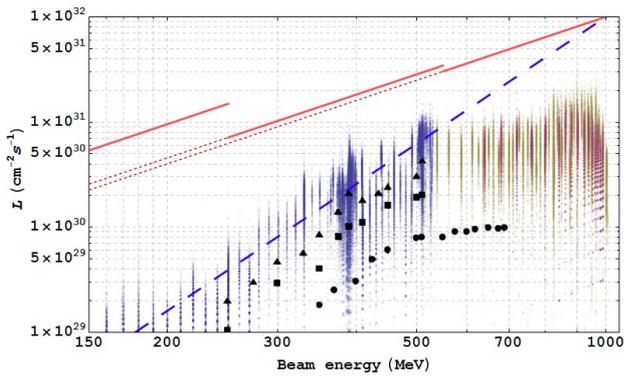

Figure 7: Luminosity scan.

In Fig. 8 the obtained beam current is presented as a function of machine operation energy. Although the current is limited not by the beam–beam effects for energies over 500 MeV but by the limited and constant positron production rate, it continues to increase with energy due to the beam's lifetime growth. The decrease of current at the highest energies is caused by the time and beam losses during energy ramping in the collider ring.

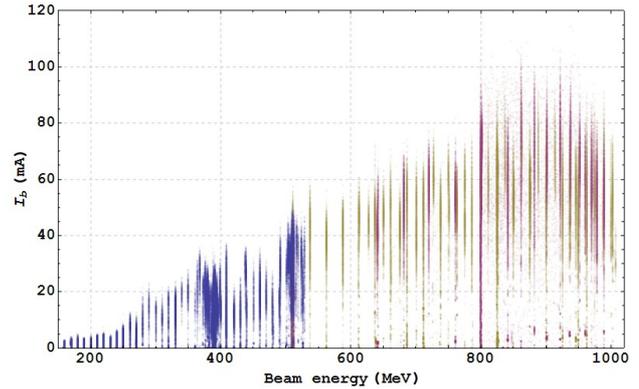

Figure 8: Beams current vs. energy.

## BEAM–BEAM EFFECTS

The real beam size can be easily obtained from the luminosity measurements. Contrary to what the simulations predict, the beam sizes grow significantly with beam current increase (see Fig. 9). However, the emittance grows monotonically, without any blow-up threshold.

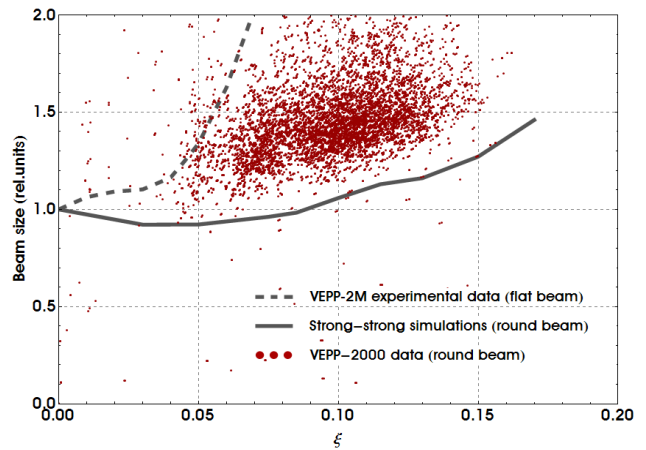

Figure 9: Beam size growth at IP ($E$ = 537 MeV).

In Fig. 9 the 'nominal' beam–beam parameter is used for the horizontal axis, which has nothing to do with a real tune shift. This parameter is a normalized measure of the beam current:

$$\xi_{\text{nom}} = \frac{N^- r_e \beta^*_{\text{nom}}}{4\pi\gamma\sigma^{*2}_{\text{nom}}}. \qquad (2)$$

## BEAM–BEAM PARAMETER EXTRACTED FROM LUMINOSITY

We can also define the 'achieved' beam–beam parameter as:

$$\xi_{\text{lumi}} = \frac{N^- r_e \beta^*_{\text{nom}}}{4\pi\gamma\sigma^{*2}_{\text{lumi}}}, \qquad (3)$$

where the beta function is nominal while the beam size is extracted from the measured luminosity. With this definition, the range of the beam–beam parameter actually achieved during experimental runs can be seen in Fig. 10.

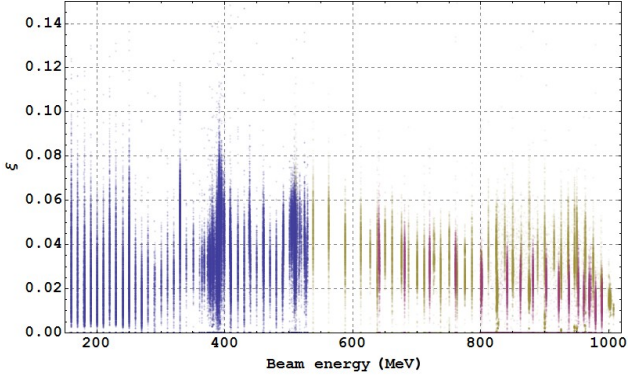

Figure 10: Achieved beam–beam parameter vs. beam energy.

The bulky data accumulated during three experimental seasons is strongly thinned out to produce Fig. 10. For this reason the top points corresponding to the peak luminosity and best-tuned machine can hardly be seen. In Figure 11 the correlation between achieved and nominal beam–beam parameters is shown for the full data at the given energy $E = 392.5$ MeV. The beam–beam parameter achieves the maximal value of $\xi \sim 0.09$ during regular work (magenta dots in Fig. 11).

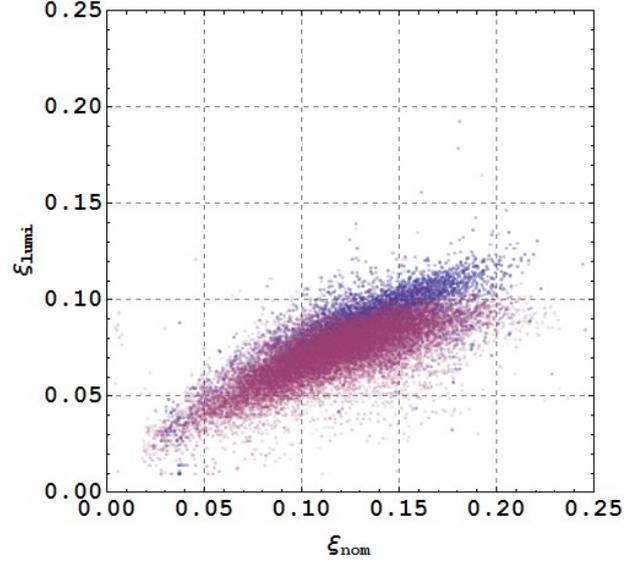

Figure 11: Achieved beam–beam parameter at 392.5 MeV.

While studying the dependence of beam–beam threshold on bunch length it was found that the RF voltage decrease from 30 kV to 17 kV gives a significant benefit in the maximal value of $\xi$ (blue dots in Fig. 11) up to $\xi \sim 0.12$ per IP. This phenomena is not yet fully explained but some predictions of beam–beam interaction mitigation can be found in Ref. [10] for the bunch slightly longer than $\beta^*$. The bunch lengthening in our particular case comes not only from the RF voltage decrease itself, but also from microwave instability, which was observed at low energies with a low RF voltage above a certain bunch intensity.

## BEAM–BEAM PARAMETER EXTRACTED FROM COHERENT OSCILLATIONS

Another independent instrument for beam–beam parameter measurement is the analysis of the coherent beam oscillation spectrum. In Fig. 12 one can find two pairs of σ- and π-modes tunes equal to 0.165 and 0.34, respectively. The total tune shift of $\Delta\nu = 0.165$ corresponds to $\xi$ per one IP equal to:

$$\xi = \frac{\cos(\pi\nu_\sigma) - \cos(\pi\nu_\pi)}{\pi\sin(\pi\nu_\sigma)} \qquad . \qquad (4)$$

The Yokoya factor here is taken to be equal to 1 due to the fact that oscillations with very small amplitude (~10 μm = 0.2 σ*) were excited by a fast kick and the spectrum was investigated for only 8000 turns. During this short time beam distribution is probably not deformed by an oscillating counter beam and remains Gaussian [11].

## FLIP-FLOP EFFECT

The beam–beam limit of $\xi_{lumi} \sim 0.1$ usually corresponds to the onset of a flip-flop effect: the self-consistent situation when one beam's sizes are blown-up while another beam's sizes are almost unperturbed. This flip-flop is probably caused by an interplay of beam–beam effects and nonlinear lattice resonances. One can see in the spectra of a slightly kicked bunch that the shifted tunes ($\pi$-mode) jumped to the 1/5 resonance in the case of a flip-flop (Fig. 13).

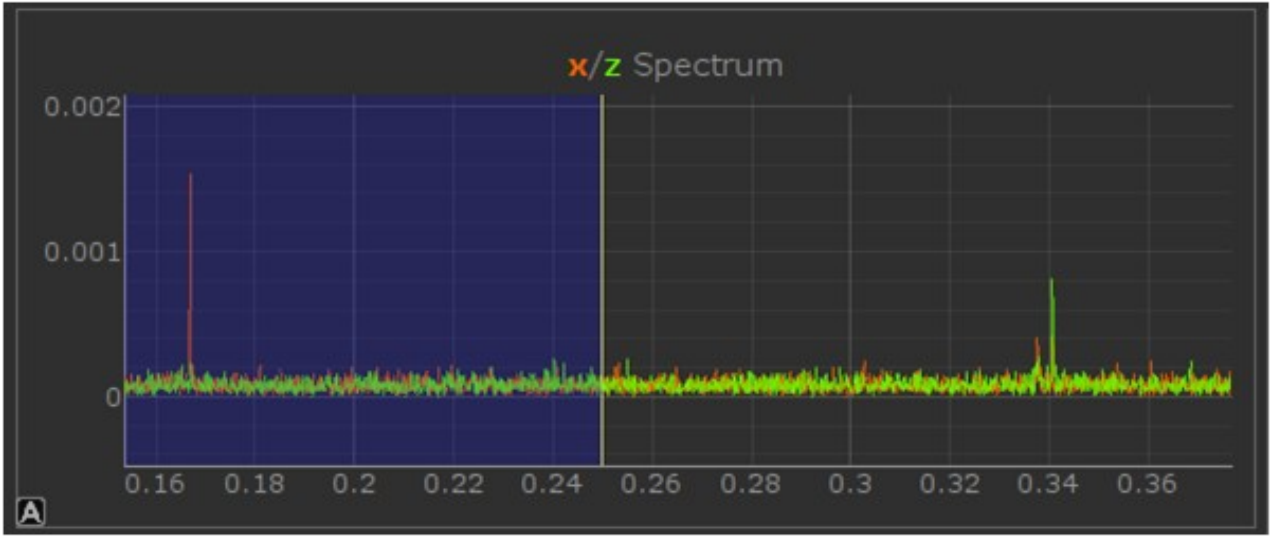

Figure 12: Coherent beam–beam oscillations spectrum at 479 MeV. The vertical axis corresponds to oscillation harmonic amplitude (mm).

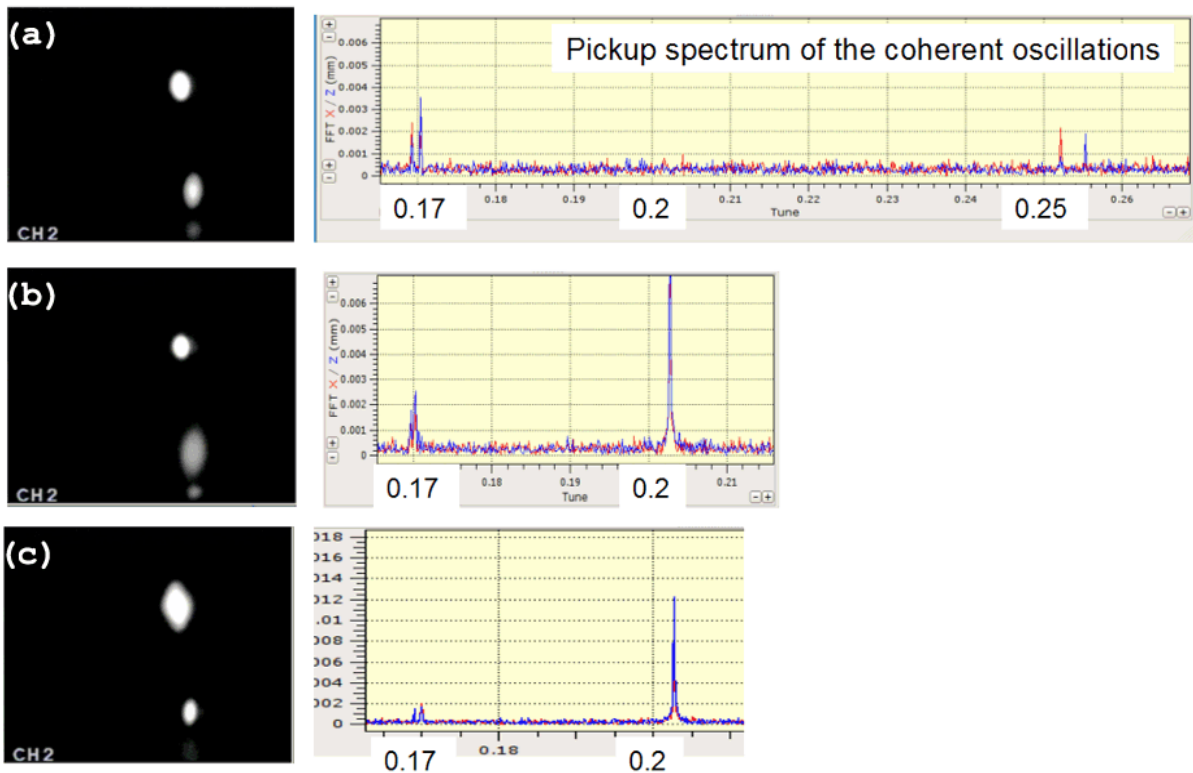

Figure 13: Flip-flop effect. 240 MeV, 5 × 5 mA. (a) Regular beams; (b) flipped electron beam; (c) positron beam.

The type of flip-flop effect that has been observed seems to be avoidable by suppressing the resonance driving terms, as well as by tuning down the working point. Unexpected problems with DA prevent us from currently using the design working point. The acceptable bunch stacking rate and beam lifetime at collision is available only for the betatron tunes of $\{v\} \sim 0.13$–$0.18$.

In Figure 13 the images from the online control TV camera are presented for the cases of regular beams, flipped electron beams or positron beams. The corresponding spectra are shown on the left.

## CONCLUSION

Round beams give a serious luminosity enhancement. The achieved beam–beam parameter value at low energy amounts to $\xi \sim 0.1$–$0.12$. VEPP-2000 is successfully taking data with two detectors across the whole designed energy range of 160–1000 MeV with a luminosity value two to five times higher than that achieved by its predecessor, VEPP-2M. To reach the target luminosity, more positrons and the upgrade of the BEP booster are needed.

## REFERENCES

[1] S. Karnaev et al. "Low Energy Luminosity at VEPP-4M Collider," Proc. APAC'01, Beijing, p. 201 (2001).

[2] Yu.M. Shatunov et al. "Project of a New Electron-Positron Collider VEPP-2000," Proc. EPAC'00, Vienna, p. 439 (2000).

[3] V.V. Danilov et al. "The Concept of Round Colliding Beams," Proc. EPAC'96, Sitges, p. 1149 (1996).

[4] L.M. Barkov et al. "Phi-Factory Project in Novosibirsk," Proc. 14th HEACC'89, Tsukuba, p. 1385 (1989).

[5] I. Nesterenko, D. Shatilov, E. Simonov, "Beam–Beam Effects Simulation for VEPP-2M With Flat and Round Beams," Proc. PAC'97, Vancouver, p. 1762 (1997).
I.A. Koop, "VEPP-2000 Project," Proc. e+e− Physics at Intermediate Energies Workshop, Stanford, p. 110 (2001).

[6] A.A. Valishev, E.A. Perevedentsev, K. Ohmi "Strong–Strong Simulation of Beam–Beam Interaction for Round Beams," Proc. PAC'03, Portland, p. 3398 (2003).

[7] D. Shatilov et al., "LIFETRAC Code for the Weak–Strong Simulation of the Beam–Beam Effects in Tevatron," Proc. PAC'05, Knoxville, USA, p. 4138 (2005).

[8] M.N. Achasov et al., "First Experience with SND Calorimeter at VEPP-2000 Collider," Nucl. Instrum. Meth. A 598 (2009) 31–32.

[9] P.M. Ivanov et al., "Luminosity and the Beam–Beam Effects on the Electron–Positron Storage Ring VEPP-2M with Superconducting Wiggler Magnet," Proc. 3rd Advanced ICFA Beam Dynamics Workshop on Beam–Beam Effects in Circular Colliders, Novosibirsk, p.26 (1989).

[10] V.V. Danilov and E.A. Perevedentsev. "Two Examples of Integrable Systems with Round Colliding Beams," Proc. PAC'1997, Vancouver, Canada, p. 1759 (1997).

[11] P.M. Ivanov et al. "Experimental Studies of Beam–Beam Effects at VEPP-2M," Proc. Workshop on Beam–Beam Effects in Circular Colliders, Fermilab, p. 36 (2001).